\documentclass[journal=jpccck]{achemso}
\setkeys{acs}{articletitle=true}
\setkeys{acs}{etalmode=truncate,maxauthors=10}
\usepackage[version=3]{mhchem} 
\usepackage[T1]{fontenc}       
\usepackage[utf8]{inputenc}
\usepackage[colorinlistoftodos,textsize=tiny]{todonotes} 
\usepackage{multirow}

\author{Sven-Hendrik Lohmann}
\affiliation{Institute of Physical Chemistry, University of Hamburg, Grindelallee 117, D-20146 Hamburg, Germany}
\author{Philip Harder}
\affiliation{Institute of Physical Chemistry, University of Hamburg, Grindelallee 117, D-20146 Hamburg, Germany}
\author{Felix Bourier}
\affiliation{Institute of Physical Chemistry, University of Hamburg, Grindelallee 117, D-20146 Hamburg, Germany}
\alsoaffiliation{Hamburg Centre for Ultrafast Imaging, Luruper Chaussee 149, D-22761 Hamburg, Germany}
\author{Christian Strelow}
\affiliation{Institute of Physical Chemistry, University of Hamburg, Grindelallee 117, D-20146 Hamburg, Germany}
\author{Alf Mews}
\affiliation{Institute of Physical Chemistry, University of Hamburg, Grindelallee 117, D-20146 Hamburg, Germany}
\alsoaffiliation{Hamburg Centre for Ultrafast Imaging, Luruper Chaussee 149, D-22761 Hamburg, Germany}
\author{Tobias Kipp}
\affiliation{Institute of Physical Chemistry, University of Hamburg, Grindelallee 117, D-20146 Hamburg, Germany}
\email{kipp@chemie.uni-hamburg.de}
\affiliation{Institute of Physical Chemistry, University of Hamburg, Grindelallee 117, D-20146 Hamburg, Germany}

\title{Influence of Interface-Driven Strain on the Spectral Diffusion Properties of Core/Shell CdSe/CdS Dot/Rod Nanoparticles}

\begin{document}

\newcommand*\mycommand[1]{\texttt{\emph{#1}}}
\newcommand{\sven}[2][]
{\todo[linecolor=red,backgroundcolor=red!25,bordercolor=red,caption={#2}, size=\tiny, #1]{\renewcommand{\baselinestretch}{0.5}\selectfont#2\par}}
\newcommand{\philip}[2][]
{\todo[linecolor=blue,backgroundcolor=blue!25,bordercolor=blue,caption={#2}, size=\tiny, #1]{\renewcommand{\baselinestretch}{0.5}\selectfont#2\par}}
\newcommand{\tobias}[2][]
{\todo[linecolor=blue,backgroundcolor=green!25,bordercolor=green,caption={#2}, size=\tiny, #1]{\renewcommand{\baselinestretch}{0.5}\selectfont#2\par}}
\newcommand{\Felix}[2][]
{\todo[linecolor=blue,backgroundcolor=yellow!25,bordercolor=yellow,caption={#2}, size=\tiny, #1]{\renewcommand{\baselinestretch}{0.5}\selectfont#2\par}}

\begin{abstract}

By combining an atomistic valence-force field approach and calculations based on the effective-mass approximation we investigate the influence of strain effects on the band alignment and general excitonic properties of core/shell CdSe/CdS dot/rod nanoparticles.
We find that the inclusion of strain effects leads to an increase in exciton energy as well as to a decrease in electron and hole wave function overlap. Importantly, the native type-I band alignment of the CdSe/CdS material system is preserved and does not change into an quasi-type-II or even type-II band offset for the nanoparticles.
Furthermore, we analyze the impact of strain on the spectral diffusion of the fluorescence emission of these nanoparticles, which is explained by migrating surface charges.
Our calculations show that the addition of strain effects leads to increased energy shifts as well as larger changes in the squared electron and hole wave function overlap, while the correlation of both also exhibits a steeper slope than for the unstrained system.
For a given CdSe core size, an increase in CdS-shell thickness decreases the possible ranges of energy shift and squared wave function overlap without changing the slope of their correlation. On the other hand, for a given nanoparticle overall thickness, dot/rod systems with a small CdSe core exhibit the strongest influenceability by surface charges.

\end{abstract}




\section{Introduction}

Semiconductor nanoparticles have shown to be advantageous in applications such as lighting and display devices.\cite{kim_full-colour_2011,yang_high-efficiency_2015} Their use in these applications is enabled by the possibility of processing them in solutions \cite{kim_contact_2008} and by the control over their bandgap-energy due to confinement effects.\cite{norris_measurement_1996} The introduction of core/shell structures not only enhances the photo-stability,\cite{hines_synthesis_1996,chilla_direct_2008} but also facilitates potential to control the emission properties in more ways than changing the emission wavelength.\cite{zhu_wave_2012}

In case of dot-in-rod particles (DRs), a spherical core material is surrounded by a rod-shaped shell material,\cite{talapin_highly_2003} effectively breaking the symmetry and allowing  for several unique optical properties: e.g. linear polarization of the emission,\cite{vezzoli_exciton_2015} higher molar extinction coefficients \cite{talapin_seeded_2007} as well as improved quantum yields \cite{talapin_highly_2003} and improved optical gain.\cite{grivas_single-mode_2013} One further notable aspect of the DR geometry depending on the material composition is the possibility to separate the charge carrier wave functions and introduce a switching mechanism like previously described for nanorod structures.\cite{rothenberg_electric_2005,muller_wave_2005} Hence, the investigation of the prevalent band offset in CdSe/CdS DRs has attracted much attention. First, a quasi-type-II alignment has been suggested, \cite{muller_monitoring_2004} later the band offset has experimentally as well as theoretically been determined to be of type-I, with the conduction-band edge of the CdSe being about 300~meV below the CdS conduction-band edge.\cite{steiner_determination_2008,wei_first-principles_2000} Nevertheless, there is spectroscopic evidence that confinement effects, determined mostly by the core size and shell geometry, can significantly influence the localization of ---in particular--- the electron wave function.\cite{sitt_multiexciton_2009} 

With respect to possible light-emitting applications, effects that negatively influence the emitter quality of core-only and core/shell nanoparticles are fluorescence intermittency, better known as "blinking", and the less prominent spectral diffusion. Both effects can only be uncovered at the single-particle level. While blinking can be nearly eliminated by the introduction of the enclosing shell and several slow-growth shell syntheses,\cite{coropceanu_slow-injection_2016} spectral diffusion is still observable in core/shell systems.\cite{muller_monitoring_2004} It has first been observed when the inhomogeneous line broadening of nanoparticle emission at cryogenic temperature was investigated within short bin times, revealing temporal energy shifts of the zero phonon line.\cite{empedocles_photoluminescence_1996,empedocles_influence_1999} Following studies have shown that the shifts are comparable to shifts induced by the quantum-confined Stark effect,\cite{empedocles_quantum-confined_1997} and that they have a discrete component behavior with an underlying memory effect.\cite{fernee_charge_2010,fernee_spontaneous_2012} 

Spectral diffusion of the light emission of nanoparticles is explained by fluctuations of the net surface charge, which interact with the electron-hole pair wave functions and thereby modify the fluorescence properties.\cite{empedocles_quantum-confined_1997,muller_monitoring_2005} We have recently investigated the spectral diffusion of large-core CdSe/CdS dot/rod nanoparticles. We have experimentally determined a correlation between the decay rate and the wavelength shift that could be modeled by migrating surface charges interacting with the photo-generated charge carrier wave functions.\cite{lohmann_surface_2017} In particular, we have shown that two elementary surface charges close to the core of these large-core DRs are sufficient to induce the changes in emission rates and wavelengths. 

The optical properties of heterostructured nanoparticles in general ---and DRs in particular--- are greatly influenced by their geometry. \cite{raino_probing_2011,sitt_multiexciton_2009,biadala_tuning_2014,wu_universal_2015}  
Changing the geometry in heterostructures does not only change the pure confinement lengths in different directions for charge carriers in the different material sections, it also changes the lattice strain distribution inside the heterostructure. Strain effects are often neglected even though their importance for basic optical properties,\cite{segarra_piezoelectric_2016,jing_insight_2015,smith_tuning_2009} particularly for the linear polarization of DR emission,\cite{vezzoli_exciton_2015,planelles_electronic_2016} has recently been uncovered.

In this paper, we report on a model to determine strain-dependent exciton properties in heterostructured nanoparticles. 
It principally allows for the investigation of nanoparticles of the whole range of material compositions, including alloys, as well as of all geometries. 
Input parameters for the modelling are lattice parameters of the unstrained materials and bulk values for conduction and valence band energies, offsets, volume-deformation coefficients, and effective masses. 
As it is experimentally not possible to switch the strain in such heterostructures on and off, a theoretical approach might be of great significance in uncovering the impact of strain.
In this work, we particularly apply the model for the CdSe/CdS core/shell DR example system, to investigate the influence of strain on the exciton properties and additionally on the spectral diffusion.
As mentioned earlier the conduction band offset in the CdSe/CdS heterostructure has been discussed actively and is rather small, thus providing an interesting example system for our investigations.
The strain distributions in DRs of different geometries are determined within an atomistic model, in which the crystal lattices of the hetero-nanostructures are being relaxed via valence force field calculations. \cite{keating_effect_1966,luo_electronic_2009,yang_strain-induced_2010,camacho_application_2010}
We implement strain-induced band edge energy changes to our EMA-based calculations. 
We first evaluate the strain-induced changes of the CdSe/CdS band alignment and of general exciton properties for different DR geometries.
Afterwards we utilize our model of migrating surface charges \cite{lohmann_surface_2017} to analyze the spectral-diffusion behavior with respect to the influence of strain.
Lastly, we investigate the geometry-dependence of the spectral diffusion behavior for several CdSe/CdS DR structures.

We find that the introduction of strain reduces the CdSe/CdS conduction band offset, but the native type-I band alignment is still retained.
In general the inclusion of strain contributions leads to an increase in exciton energy and a decrease in squared wavefunction overlap. 
The investigation of spectral diffusion shows that the strain-induced lowered conduction-band offset leads to a stronger influenceability of the charge carrier wave functions by migrating surface charges.
Consequently, the ranges for energetic shifts and relative squared wave function overlaps are both increased.
Regarding the geometry-dependence of spectral diffusion we find that decreasing the shell thickness, for a set core diameter, causes a greater magnitude in exciton-energy and wave function overlap changes.
For a fixed shell geometry the charge carrier wave function can most strongly be influenced by the surface charges for very small core sizes.
With this knowledge the DR geometry can be adjusted to minimize the effects of spectral diffusion, which diminish the emission quality of these particles.

\section{Description of the Model}

The theoretical modeling relies on two main ingredients: (i) The calculation of strain distributions in hetero-nanostructured DRs with their subsequent translation into potential landscapes for electrons and holes and (ii) the calculation of electron and hole wave functions and energies in arbitrarily-shaped potentials within the effective-mass approximation (EMA).
The latter ingredient relies on solving the Schrödinger equation on a three-dimensional spatial grid. The method is based on a one-dimensional approach, described in Ref. \citenum{strelow_light_2012}, which has been expanded to a three-dimensional cartesian grid and modified into a two-band EMA-based calculation for semiconductor nanostructures for the investigation of excitonic complexes in CdSe nanowires.\cite{franz_quantum-confined_2014}
The calculations can be performed in a self-consistent way, by iteratively computing the wave function of an electron in the potential of a hole and vice versa, such that the direct Coulomb interaction between electrons and holes is taken into account.
The two-particle exciton energy is calculated by adding the bulk-bandgap energy and the determined single-particle electron and hole energies and, since both single-particle energies contain the mutual Coulomb energy, by subtracting this Coulomb energy once.
It is the free variability of the external potential that allows for the implementation of Coulombic point charges located on the DR surface \cite{lohmann_surface_2017} or --as will be presented in this paper-- the effect of local strain distributions.

\begin{figure*}[htbp]
  \includegraphics{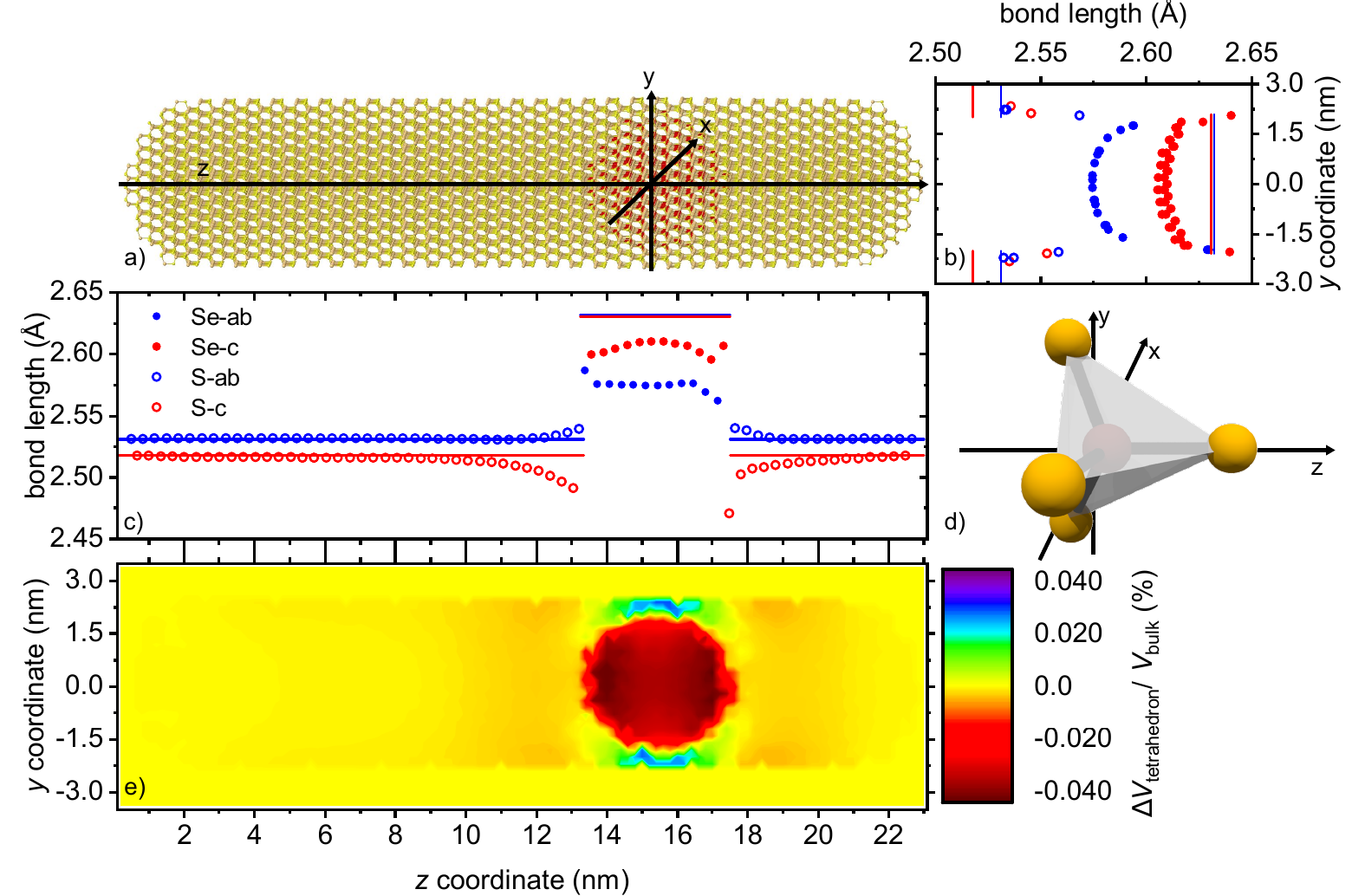}
    \caption{(a) Atomistic model of a DR with a 3.8 nm CdSe core enclosed in a CdS shell of a total length of 22 nm and thickness of 5.1 nm. The wurtzite [001] direction is parallel to the $z$-axis. (b,c) Bond-length profiles along the $y$- and $z$-axis, respectively. Solid lines give the bulk bond lengths while data points represent the calculated bond lengths of the strain-relaxed structure. Filled and open data points  correspond to the Se-Cd an S-Cd bonds, respectively. Blue and red color coding corresponds to $ab$- and $c$-bonds, respectively. (d) Sketch of a tetrahedron (either Cd$_4$S or Cd$_4$Se) that defines the elemental volume element of the DR. So-called $c$-bonds are aligned parallel to the $z$-axis, $ab$-bonds have large components in the $xy$-plane. (e) Relative volume change due to strain relaxation for the $yz$-plane depicted as a two-dimensional false-color plot.}
\label{figure1}
\end{figure*}

Here, we would like to mention that we are aware of the limitations of our EMA-based two-band model. Using more sophisticated calculations, such as DFT or pseudo-potential based ones, would offer a deeper insight to the exciton properties, disclosing information on the exciton fine-structure level. Instead, we have opted for this robust and quick method in order to be relatively independent of the nano structure geometry/size. This way it is possible to screen a broad range of geometries to determine their effect on the basic exciton properties. Furthermore, using this approach the geometry variation is easily combined with other effects such as migrating surface charges or strain, while the simplicity of our model allows us to retain manageable computing times.

In order to incorporate strain effects into our calculations, we no longer use a simplified step-wise box potential to model the hetero-nanostructure systems, but shift to an atomistic approach. 
Atomistic models of the CdSe/CdS DR particles are being generated as follows.
First, a pure wurtzite CdS crystal is defined with bulk parameters for bond lengths and angles. The crystal consists of Cd$_4$S tetrahedrons, one of which is depicted in Fig.\ \ref{figure1} (d) to illustrate the two different bond types of the wurtzite crystal structure. Every sulphur atom is tetrahedraly bound to four cadmium atoms, one of which is directly in the $z$-direction and the other three exhibit components in the $xy$-directions. 
The difference between bonds with $xy$-components and bonds in $z$-direction is taken into account by their respective parametrization.
In the following we will refer to bonds with large components in the $xy$-plane as  $ab$-bonds and bonds in $z$-direction as $c$-bonds, according to their crystallographic labeling.
The rod-like shape is then obtained by cutting the CdS crystal into the intended geometry.
After that, the CdSe core of a certain diameter and at a certain position is being defined by replacing the sulphur atoms with selenium atoms in the specified area. By keeping the atomic positions fixed, the emerging atomistic model can be regarded as a highly strained CdSe dot, since the CdSe bonds are defined by CdS bulk values, embedded in an unstrained rod-shaped CdS shell. 
In a next step, a valence force field proposed by Keating et al.\cite{keating_effect_1966,camacho_application_2010} has been utilized for the relaxation of the atomistic models.
The total energy of the system is minimized using a combination of hill climbing and simulated annealing algorithms. Open boundary conditions were applied and therefore the possible influences of surface ligands on the structures were not incorporated into the model. By variation of both bond lengths and angles, optimized CdSe/CdS DR structures were obtained, one of which is exemplarily shown in Fig.\ \ref{figure1} (a).

Figure \ref{figure1} (b) depicts the bond-length profiles along the $y$-axis, comparing the lengths of $ab$-bonds (blue data points) and $c$-bonds (red data points) in our atomistic model to the respective bulk bond lengths (solid lines). Obviously, along this direction, both $ab$- and $c$-bonds in the CdSe core are compressed, whereas the CdS bonds in the shell region are stretched. 
The center region of the CdSe core is compressed stronger than in areas near the interface.
At the interface the compressive strain is efficiently relieved into the CdS shell.
The deformation of the CdSe $ab$-bonds is larger than of the CdSe $c$-bonds. The $ab$-bonds posses components of all three spatial directions instead of only one, consequently the acting force is higher in comparison to $c$-bonds.

The bond-length profiles along the $z$-axis, shown in Fig.\ \ref{figure1} (c), reveal a similar behavior with respect to the compression of the core region and the stronger bond length deformation of the $ab$-bonds in this region.
Differences can be observed for the distribution of strain in the shell region. Here, only the CdS $ab$-bonds are expanded, while the $c$-bonds are compressed.  
Originating from the anisotropic geometry, there is a much thicker shell in $z$-direction than in the $xy$-plane, leading to a much more rigid backbone of the shell in the $z$-direction.
The counterforce of the compression of the core in $z$-direction leads to a compression of the CdS $c$-bonds at the core/shell interface. 
On the other hand, the counterforce of the core compression leads to an expansion of the CdS $ab$-bonds at the interface, similar as discussed above for the $y$-direction.

After the position-dependent strain information has been obtained, this data is transformed into a spatial potential landscape, i.e., into position-dependent strain-induced shifts of the band edge energies.
This can be done by the utilization of volume-deformation coefficients.\cite{jing_insight_2015}
For that it is necessary to quantify the volume change in each volume element.
The volumes of all Cd$_4$Se and Cd$_4$S tetrahedrons are calculated for the relaxed atomistic CdSe/CdS DR structure.
Then, for each tetrahedron, the volume change $\Delta V$ is determined by comparison to the volume of tetrahedrons with bulk parameters ($V_{bulk}$).
Fig.\ \ref{figure1} (e) shows a two-dimensional false-color plot of the tetrahedron-volume change in the $yz$-plane, revealing compressed volume elements inside the core and both compressed and extended volume elements inside the shell close to the core/shell interface.
The change in band edge energies for each volume element can now be calculated as $\Delta E =\alpha _{V} \ln (1- \Delta V / V_{bulk})$, with $\alpha_V$ being the volume deformation coefficient for the conduction and valence band, respectively.\cite{smith_semiconductor_2010}

In the last step our atomistic model structure has to be converted into external potential landscapes for electrons and holes, suitable as input for our EMA-based calculations.
For that a regular cartesian three-dimensional grid is defined and overlaid with the atomistic structure consisting of strained tetrahedrons. 
To each unit cube of the grid that contains a center coordinate of a Cd$_4$Se or Cd$_4$S tetrahedron, corresponding specific material constants ---such as valence and conduction band levels, effective charge carrier masses and relative permittivities---  are assigned. The material constants for the other unit cubes are set by interpolation. Such a mapping of the atomistic model onto a cartesian grid allows for the calculation of strain effects on the electron and hole wave functions and energies in hetero-nanostructures within our three-dimensional EMA-based model.
At this point we would like to point out that the wave function we refer to is the envelope wave function in the framework of the EMA-based two-band model. The parameters used in the atomistic model, the valence force field approach, and the EMA-based  calculations are given in Table \ref{table:parameters}.

\begin{table}
\caption{Input parameters for both the valence force field and effective mass based calculations for the different materials}
\label{table:parameters}
\begin{tabular}{ l c c }
\hline \hline
Parameter										& CdSe & CdS \\ 
\hline 
$ab$-bond length	(\AA)							& 2.6321  & 2.5310 \\
$c$-bond length	(\AA)							& 2.6307  & 2.5179  \\
bond elasticity	(N/m$^2$)						& 22.878\cite{madelung_semiconductors_2004}  & 25.029\cite{bolef_elastic_1960} \\
bandgap energy (eV)									& 1.74  & 2.42   \\
CdSe conduction band offset (eV)	& \multicolumn{2}{c}{-0.30}	\\
effective electron mass						& 0.13\cite{haus_quantum_1993} & 0.20\cite{mews_preparation_1994}  \\
effective hole mass							& 0.45\cite{haus_quantum_1993} & 0.70\cite{mews_preparation_1994}  \\
$\alpha_V$ valence band \cite{wei_predicted_1999} (eV)		& -1.81 & -1.51  \\
$\alpha_V$ conduction band\cite{wei_predicted_1999}	 (eV)	& -3.77 & -3.59  \\
\hline \hline
\end{tabular}
\end{table}

In the following we will often compare results for DRs obtained, on the one hand side, without taking strain into account, and, on the other hand side, with taking strain into account. We will call it the unstrained case, when the atomistic modeling has not been used and just the simple step-wise box potential of core/shell DRs is applied for the calculations of excitonic properties. In contrast, in the strained case, the modified potential deduced from the strain-relaxed atomistic model is utilized.

\section{Results}
\subsection{Influence of Strain on Uncharged CdSe/CdS DRs}
\subsubsection{Band Alignment at the CdSe/CdS Interface}

\begin{figure}
  \includegraphics{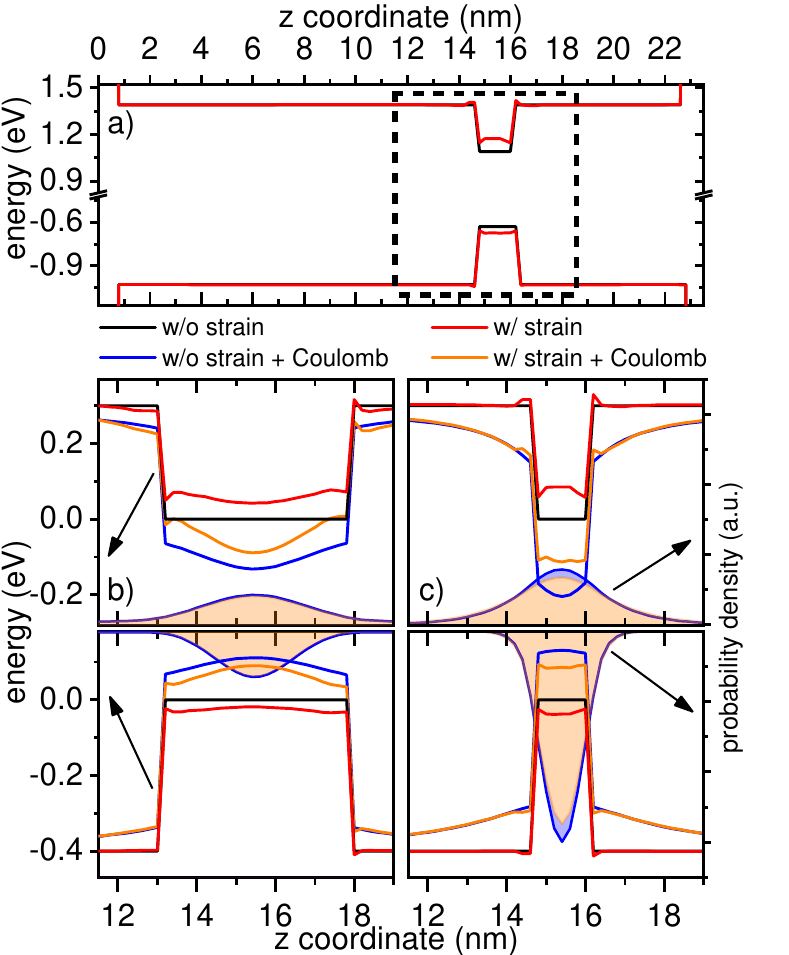}
    \caption{(a) Valence and conduction band energies along $z$ for a simulated CdSe/CdS DR particle (1.8 nm CdSe core diameter, 22 nm CdS shell length and 5.1 nm shell diameter) without/with (black/red line)  strain contributions. The dotted square marks the region of special interest.  (b, c) Sections of the valence and conduction bands (colored lines) and probability densities of the hole and electron ground states (curves with their area below filled) along $z$ in the core region of DRs with the same shell dimensions and core diameters of (b) 4.8 nm and (c) 1.8 nm. 
Black and red lines correspond to bands like in (a), where no Coulomb interaction is taken into account. Blue and orange lines have been self-consistently calculated taking the direct electron-hole Coulomb interaction into account, without and with contributions of strain effects, respectively. These latter bands are the basis of probability densities depicted in (b) and (c). }
\label{figure2}
\end{figure}

The band alignment of the CdSe/CdS DR structures has been a controversially discussed topic since their first successful synthesis. Spectroscopic data has suggested a possible quasi-type-II band offset.\cite{muller_monitoring_2004}
Later theoretical calculations and experimental data have confirmed a type-I alignment, with the CdSe conduction-band edge being 300~meV below the CdS band edge,\cite{steiner_determination_2008,wei_predicted_1999} but the effective charge carrier localization being dependent on the CdSe core size.\cite{sitt_multiexciton_2009}
Here, we investigate the impact of strain on the band alignment in the CdSe/CdS DR system and conduct a comparison for different core sizes.
Figure \ref{figure2} (a) portrays the changes of the band-edge energy levels of a CdSe/CdS DR in a profile along the rotational symmetry axis.
The black line represents the energy level for the DR without strain contribution (unstrained system).
The red line depicts the DR band energy levels under consideration of the strain-related band-edge energy changes deduced from the strain-relaxed atomistic models (strained system).
Since Fig.\ \ref{figure1} shows the largest bond-length deviations are in the core region, the strain induced energy changes are also concentrated in the core region highlighted by the dotted square in Fig.\ \ref{figure2} (a).
For this reason we will take a closer look at the core region for CdSe cores of two different sizes embedded in the same shell geometry, the CdSe cores being 4.8 nm and 1.8 nm in diameter.

We first examine the structure with 4.8 nm core diameter [Fig.\ \ref{figure2} (b)] and concentrate on the influence of strain effects as well as Coulomb interactions on the band-edge energies. Evaluating first purely the effect of strain [same color code as in Fig.\ \ref{figure2} (a)] demonstrates an energy increase for both the valence and conduction band in the CdSe core. 
The difference in magnitude of the energy level change can be attributed to the different volume-deformation coefficients for valence and conduction band respectively.
The band edge energy change is point-symmetric to the center of the CdSe core with the minimum being at this position.
The impact of strain on the band-edge energy in the shell region is negligible.
Now we want to evaluate the impact of Coulomb interaction between electron and hole, which is treated by self-consistently calculating the Coulomb potential for one charge carrier in the presence of the other.
The blue line in Fig.\ \ref{figure2} (b) represents the emerging potentials of the unstrained DR structure.
Note that the Coulomb interaction itself is a many-body problem, which is, in this case, visualized in a single-particle picture in one spatial dimension. 
The consideration of Coulomb interaction leads to an energy decrease.
The comparison of the influence from strain (red line) and Coulomb (blue line) contributions on the band-edge energy illustrates their opposed effects.
Considering both strain and Coulomb effects lead to the orange lines for the conduction and valence bands in Fig.\ \ref{figure2} (b), essentially representing the combination of the features of the red and blue lines.
If one compares the orange lines with the blue ones, i.e. the bands under consideration of Coulomb interaction with and without taking strain into account, it can be seen that strain effects generally decrease the band offsets between the core and the shell region.
The changed band offsets should also lead to changed electron and hole wave functions. 
Correspondingly, in Fig.\ \ref{figure2} (b) the probability density of the respective charge carrier in $z$-direction are depicted as curves with their area colored accordingly.
The probability densities of electron and hole nearly coincide for the unstrained and strained system, hence, in the depiction of Fig.\ \ref{figure2} (b), they cannot be discriminated. A closer look into the data reveals that the probability density of the strained system is slightly smaller in the core region than for the unstrained system, which is consistent with the decreased band offset.

For CdSe cores of 1.8~nm in diameter [Fig.\ \ref{figure2} (c)] we also observe a large change in the CdSe band edge energy due to compression of the CdSe bonds when comparing the strained structure (red line) with the non-strained one (black line).
Similar to the 4.8 nm diameter CdSe core the change in band-edge energies is also symmetrical, but the maximum increase in band-edge energy is in the core center.
The smaller cores are completely strained and thus do not exhibit a local energetic minimum along the $z$-axis.
Coulomb interaction leads to a decrease of the energy bands. The relatively flat energy band of the orange curve hints at a compensation of the Coulombic interaction by the strain contribution.
In the shell region the band-edge energy is only slightly elevated for the small cores, which is consistent with the compression of the CdS bonds in $z$-direction. 
Differences of the probability densities for electron and hole between the unstrained and strained system are slightly larger than in the case the DR with the 4.8 nm CdS core. These differences can now be seen by the small blue filled area behind the orange-filled area in Fig.\ \ref{figure2} (c), the latter representing the curve for the strained system. Strain leads to a slightly larger leakage of the probability density of electrons and holes into the shell region.

Overall it can be said that the increase in band-edge energy through strain is slightly higher for smaller cores, but this is not enough to change the 300~meV type-I band offset of the CdSe/CdS heterostructure into a quasi-type-II or even type-II band alignment.
The spatial probability density of the charge carrier wave functions is for both the electron and hole highest in the center of the core region.
Because the rather large valence-band offset induces a tight confinement, the hole wave function is very strongly localized in the CdSe core, even for small cores.
The electron wave function is also bound to the core region, but because of the smaller conduction band offset and effective mass it is slightly protruding into the surrounding CdS shell.
The addition of strain lowers the probability density in the core region for both charge carriers.
Because of the three-dimensionality of the problem, the overall change in wave function overlap can not directly be determined from one-dimensional profiles like shown in Fig.\ \ref{figure2} (b) or (c). Instead it will be discussed in the following section in more detail.

\subsubsection{Exciton Properties}

\begin{figure}[htbp]
  \includegraphics{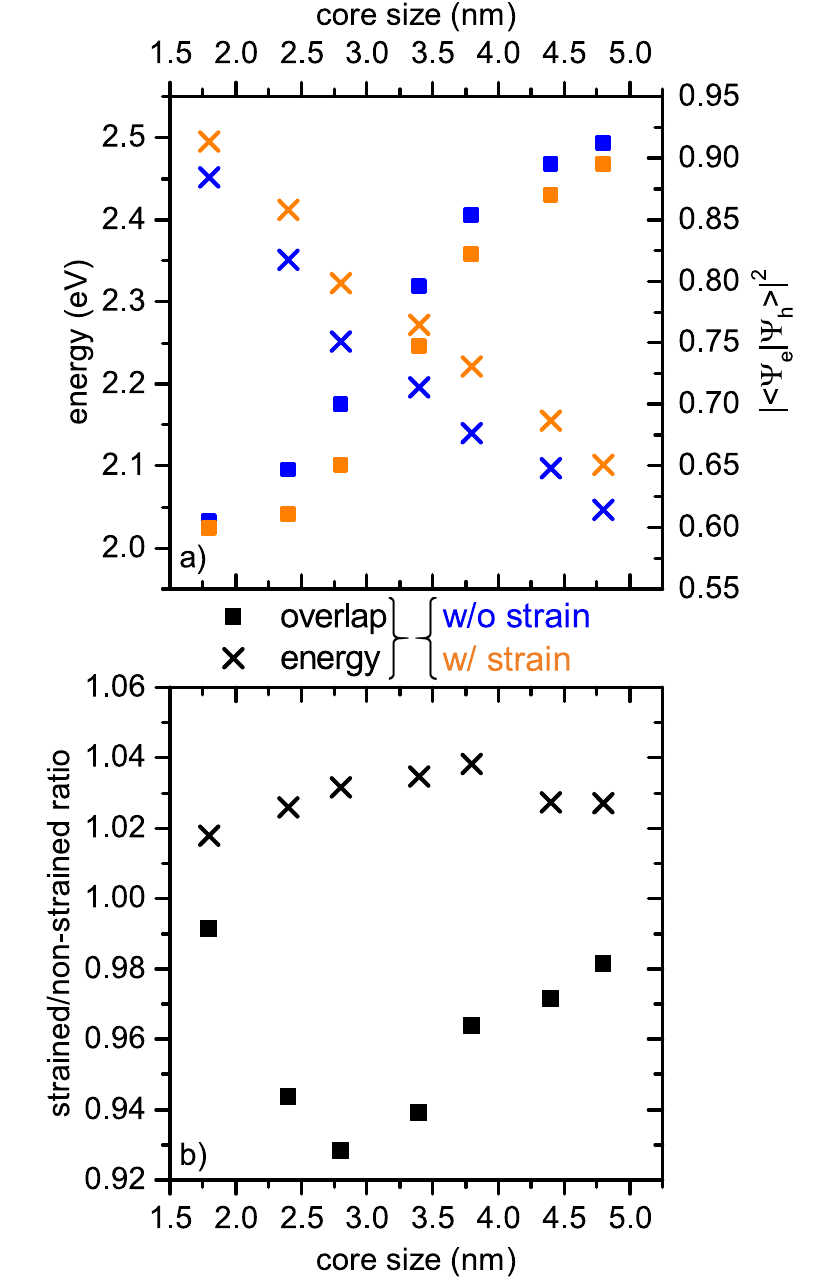}
    \caption{(a) CdSe core size dependence of the exciton energy (crosses) and squared wave function overlap (squares) for DR particles of the same CdS shell dimensions (22 nm length and 5.1 nm thickness) without/with strain contribution (blue/orange). (b) Core size dependence of the ratios between the exciton energy (crosses) and squared wave function overlap (squares) of the strained and the unstrained CdSe/CdS DR structure.}
\label{figure3}
\end{figure}

In Fig.\ \ref{figure3} (a) the exciton energy (crosses) and squared wave function overlap (squares) are plotted in dependence of the core size of DR structures with the same CdS shell geometry, for both the unstrained (blue) and strained (orange) case. 
Here, and in all following discussions, Coulomb interaction is taken into account in a way that the two-particle nature of the excitionic direct Coulomb interaction is correctly considered (see Section II). Furthermore, we would like to remind at this point that the squared overlap of the envelope functions is considered.
Independent from whether strain has been taken into account or not, with increasing CdSe core size, the exciton energy declines approximately linear and with a similar slope.
The energy of the strained DR structures is on average shifted by 50 meV to higher energies.
The similar slope is a consequence of a complex interplay between confinement energy, Coulomb interaction, and strain contributions.

In contrast to the exciton energy, the squared wave function overlap in dependence of the core size displays a sigmoid behavior. Compared to the unstrained model, the strained structures show a smaller squared wave function overlap,
with the differences between both model structures being more pronounced for intermediate core sizes.
Since, for the strained structure, the hole wave function is still tightly confined to the core region we can deduce that strain effects lead to a higher delocalization of the electron wave function.

For small core sizes, the squared wave function overlap converges to a minimal value around 60\%. Here, the calculated exciton energy is about 2.49 eV, larger than the bulk bandgap of CdS.
Thus, a strong delocalization of at least the electron wave function is generally to be expected.
We note that an exciton energy above the CdS bandgap does not necessarily mean that the charge carrier wave functions are wide-spread into the CdS shell. This is because of the anisotropic geometry and the resulting confinement potential in different spatial directions.
In the $xy$-direction the core is surrounded by a significantly thinner CdS shell than in $z$-direction. 
The shell anisotropy leads to a higher confinement-energy contribution in $xy$-direction and therefore to a different wave function delocalization than in $z$-direction.
In a three-dimensional heterostructure, such as a CdSe/CdS DR particle, distinct combinations of core size, shell thickness, and length can induce cases, where the overall electron energy is higher than the band offset, but the wave function localization in $z$-direction still prevails.
The confinement in $xy$-direction lifts the exciton energy far above the band offset. In $z$-direction however, because of the thicker shell, the electron energy remains below its confining potential preventing the delocalization in this direction. 
The comparatively large squared wave function overlap calculated here is an indicator for the retained type-I band offset. For comparison, calculations for a DR structure with the same geometry but an artificial quasi-type-II band offset between CdSe and CdS show a considerably lower squared wave function overlap of only 0.23 and thus a stronger spatial separation of electron and hole.

To further analyze the effect of strain on the exciton properties,
the exciton energy and squared wave function overlap of the strained DR structures are normalized to the values of their unstrained counterparts.
Figure \ref{figure3} (b) shows these normalized values in dependence of the CdSe core size.
Taking strain into account increases the exciton energy (crosses) for all core sizes.
The minimum energy increase is 1.8 \% for the smallest core size.
The maximum change in energy occurs at a core size of 3.8 nm with an increase of 3.8\%.
The squared wave function overlap (squares) is reduced for all core sizes varying between changes of 2.85 and 7.3 \%.
In contrast to the exciton energy the maximum change in squared wave function overlap occurs at a core size of 2.8 nm.

The minimum at this specific core size is interesting because it occurs at the same core size for which Sitt et al.\ observed a cross-over from  attractive biexciton-binding energies for larger cores to repulsive biexciton-binding energies for smaller cores in a previous work on CdSe/CdS DRs with varying core sizes.\cite{sitt_multiexciton_2009}
The authors labeled this cross-over as a transition from a type-I charge carrier localization to a quasi-type-II charge carrier delocalization, which refers to a change in the spatial localization of the electron. 
Here, it is  important to note that this transition does not relate to the type of band offset itself, which is still assumed to be type-I.\cite{sitt_multiexciton_2009}
In a simple picture, the idea behind this cross-over is that upon decreasing the core size, the confinement energy is increased such that eventually the $z$-component of the confinement energy is close to the conduction-band offset between CdSe and CdS. Consequently the related wave function to some extent protrudes into the shell. 
The effect of delocalization at this core size is much larger for biexcitons. It leads to a transition from binding biexciton energies, as typically observed in type-I heterostructures to anti-binding biexciton energies, as observed in actual type-II structures.

We can explain the minimum of the ratio between the squared wave function overlap of the strained and the unstrained DR, shown in Fig.\ \ref{figure3} (b), with a similar reasoning. 
As we have elaborated above, taking strain into account essentially decreases the confinement potential in every spatial direction by lowering the conduction band offset.
Starting from the unstrained DR, for large core diameters the confinement energies are relatively low, so that the change in conduction-band offset by strain effects is not enough to overcome the energetic barrier between core and shell. 
For very small core diameters the $z$-component of the confinement energy is already as large as the band offset, thus, the strain-induced further reduction of the conduction-band offset does not show a big impact on the relative wave function overlap.
For an intermediate core size, when the $z$-component of the confinement energy is close to the conduction-band offset, the strain-induced decrease in conduction-band offset can lead to the overcoming of the confinement potential in $z$-direction and thus to more significant delocalisation of the electron wave function and, consequently, to a more distinct decrease in the exciton squared wave function overlap. This behavior can be observed in Fig.\ \ref{figure3} (b) for a core size of 2.8 nm.

In light of the above discussion, we want to stress again that the native type-I band offset is preserved for all core sizes also when strain is considered.
One might describe the situation in DRs with small cores by a quasi-type-II charge carrier delocalization, but even then a large part of the electron wave function remains localized in the core region because of the Coulomb interaction between electron and hole.

\subsection{Migrating Surface Charges on CdSe/CdS DRs}
\subsubsection{Influence of Strain on the Spectral Diffusion}

In this section, we will analyze how the previously discussed strain-induced changes of the band structure influence the characteristics of spectral diffusion.
We have already shown that fluctuating surface charges can induce spectral diffusion and, here, we combine our spectral-diffusion model\cite{lohmann_surface_2017} with above described strain effects and calculate the exciton properties.
The changes in exciton energy and squared wave function allow us to predict the influence of surface charges on the emission wavelength as well as the decay rate.
\begin{figure*}[htbp]
  \includegraphics{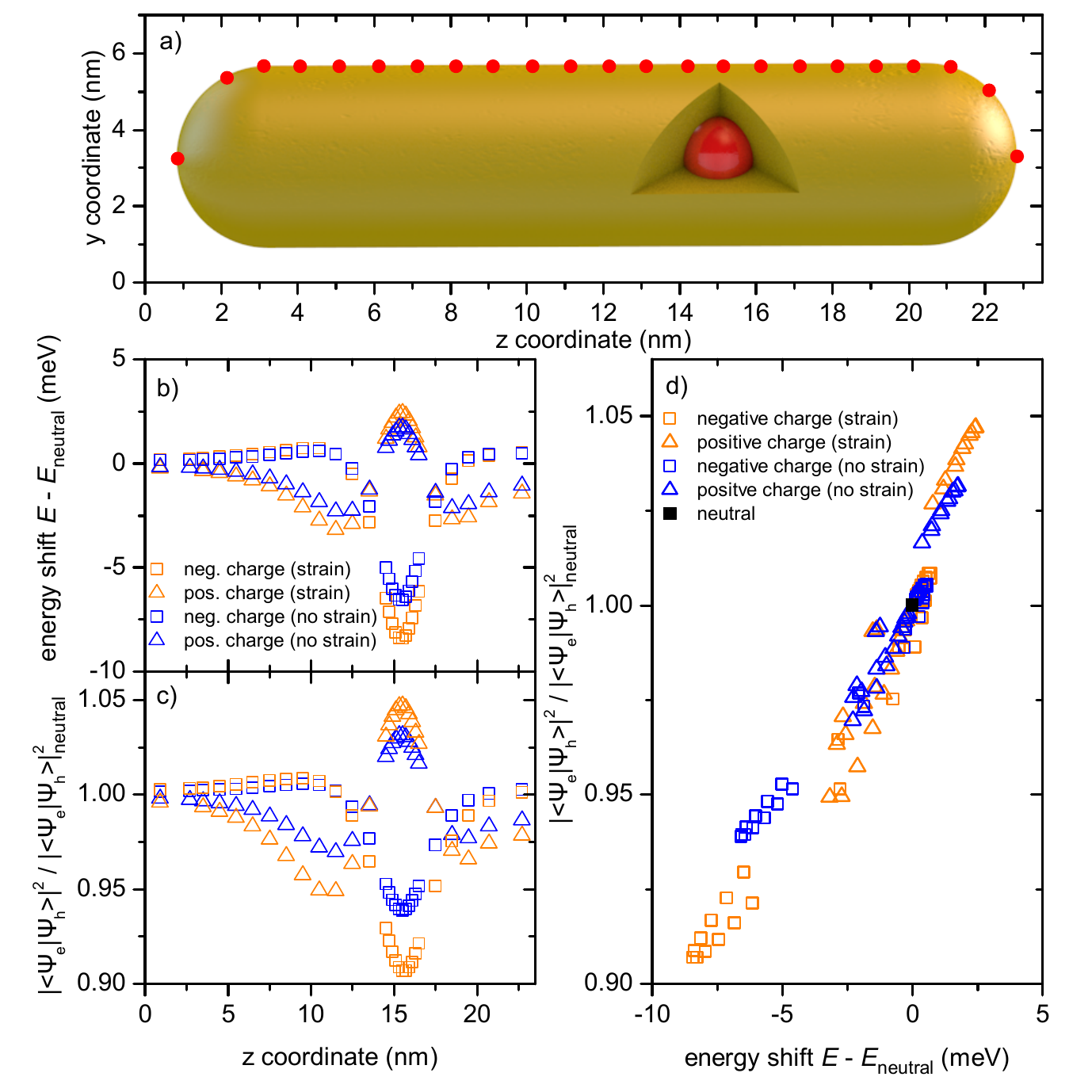}
    \caption{(a) Sketch of a CdSe/CdS DR nanoparticle with possible surface charge positions marked as red dots. (b) Surface-charge induced shift of the exciton energy in dependence of the surface-charge's $z$-position. (c) Relative squared wave function overlap of electron and hole in dependence of the surface-charge's $z$-position. Reference point for (b) and (c) are the values for the neutral CdSe/CdS DR structure, i.e. without a surface charge. (d) Correlation of the relative squared wave function overlap and the energy shift. In (b-d), squares and triangles represent calculations for a negative and a positive surface charge, respectively. Blue and orange markers represent calculations without and with taking strain into account, respectively.}
\label{figure4}
\end{figure*}

In order to model the surface-charge migration, individual Coulombic point charges are placed on the DR surface and the exciton properties have been calculated for different positions of these charges.
We have previously described the general effect of surface charges on the exciton properties for CdSe/CdS DRs with a rather large core diameter of 4.7 nm.\cite{lohmann_surface_2017}
For that system, only surface charges in direct proximity to the CdSe core significantly influence the exciton properties, with negative surface charges generally inducing a stronger effect than positive ones. 
Here, we first consider a DR with a small core, as sketched in Fig.\ \ref{figure4} (a).
The CdSe core has a diameter of 1.8 nm, the CdS shell has dimensions of 22 nm in length and 5.1 nm in diameter.
Individual Coulombic charges are placed on the rod surface at positions every 1 nm (0.2 nm near to the core) along the $z$-coordinate, like marked with red dots in Fig.\ \ref{figure4} (a), thereby exploiting the rotational symmetry along this axis.

Figure \ref{figure4} (b) gives the surface-charge induced shift of the exciton energy and (c) shows the relative squared wave function overlap, both in dependence of the surface-charge position.
It is distinguished between positive and negative surface charges (squares and triangles, respectively). Blue data points correspond to results obtained for the unstrained model structure, while orange data point represent the model in which strain effects are included. 

The general trends for both, the unstrained and strained model systems, curves can qualitatively be explained as follows. 
Regarding first a positive surface charge, at a position far away from the core there is nearly no interaction and no effect on both energy and overlap. Moving the positive surface charge in direction of the core first leads to a decrease of energy and overlap. This is because the surface charge starts to interact first with the more delocalized electron while the hole remains unaffected. Attractive interaction between surface charge and electron decreases the overall energy. At the same time, the electron wave function is drawn in direction of the surface charge and thus the electron-hole overlap is decreased.
Moving the positive surface charge further to the center position of the core finally leads to an increase of energy and overlap. This is because at some point also the repulsive interaction of the positive surface charge with the hole becomes important. Because of the larger effective mass of the hole, this repulsion ultimately exceeds the attraction between surface charge and electron, leading to an increase in energy. At the same time the more delocalized electron wave function is drawn into the core region and thus the overlap is increased. Regarding now a negative surface charge, the movement from far away towards the position of the core yields inverse trends for energy and overlap, which can qualitatively be explained like above.


Addressing now the effect of strain, we generally observe increased influence of surface charges on both the exciton energy shift and the squared wave function overlap when strain is considered, as can be seen by the larger vertical spreading of orange data points compared to blue data points in Fig.\ \ref{figure4} (b) and (c), respectively.

Going into details for the squared wave function overlap first [see Fig.\ \ref{figure4} (c)], the strongest effect of strain occurs for surface charges in direct proximity to the CdSe core.
Negative surface charges in the core region lead to a stronger decrease in squared wave function overlap if the DR structure is strained.
Similarly, strain effects induce an increase in squared wave function overlap for positive surface charges placed in the core region.
Furthermore positive surface charges situated outside the core region decrease the squared wave function overlap stronger when strain effects are considered.
In order to explain these features, we will shortly recapitulate the influence of strain:
The band offsets are lowered by the inclusion of strain effects.
Because of the higher volume expansion coefficients, the change is more pronounced for the conduction band.
Combined with the inherently smaller offset of the conduction band, the electron wave function is delocalized stronger than the hole wave function.
The stronger delocalization of the electron wave function corresponds to a larger influence of external effects such as surface charges.
Therefore in the model structure with strain contributions, the electron wave function is pushed farther away by negative surface charges, while being attracted more strongly by positive surface charges.

The exciton energy in dependence of the charge position in $z$-direction, shown in Fig.\ \ref{figure4} (b), exhibits a similar behavior as the squared wave function overlap regarding the differences between the unstrained and strained structure. 
Including strain into the model increases the effect of surface charges.
The specific characteristics regarding the exciton-energy changes between the unstrained and strained DR model systems originate from a complex interplay of confinement and Coulomb potential alterations as well as charge carrier effective masses and mobilities.
Thus we will not specifically discuss every difference in exciton energy between both model structures, but give a more general explanation of the changes.
Since electrons are delocalized more strongly, here, the spatial range over which interaction with surface charges occur is larger than for holes.
Therefore the energy change induced by surface charges outside the core region is dominated by their interaction with the electron, even more so for the strained structure.
For surface charges in the core region, the exciton energy is dominated by the interaction between surface charge and the hole, because of its higher effective mass compared to the electron.

Figure \ref{figure4} (d) summarizes the influence of strain on the spectral diffusion in CdSe/CdS DRs. 
It combines the plots of Fig.\ \ref{figure4} (b) and (c) into a correlation between the change in normalized squared wave function overlap and shift of the  exciton energy.
This correlation is of particular interest because it is experimentally accessible via the measurement of spectral diffusion. 
Both the unstrained and strained model system reveal a similar trend: a red-shift in exciton energy comes along with a decrease in squared wave function overlap.
The strained model system exhibits expanded ranges for both the wave function overlap and the exciton energy, as a consequence of the more pronounced extrema discussed with respect to Fig.\ \ref{figure4} (b) and (c).
Furthermore the trend for the strained structure exhibits an overall steeper slope than the unstrained one, because~--~as noticeable in the position-dependent plots in Fig.\ \ref{figure4} (b) and (c)~--~the relative changes induced by the surface charges are stronger for the squared wave function overlap than for the exciton energy.

\subsubsection{Impact of the Geometry on the Spectral Diffusion for Strained DRs}

In this section, we investigate the dependence of the spectral-diffusion properties on the geometry of CdSe/CdS DRs, for which strain effects are taken into account. We are thus combining the model of randomly diffusing surface charges with the strain model for DRs with different core and shell dimensions and calculate the exciton properties.

\begin{figure}
  \includegraphics{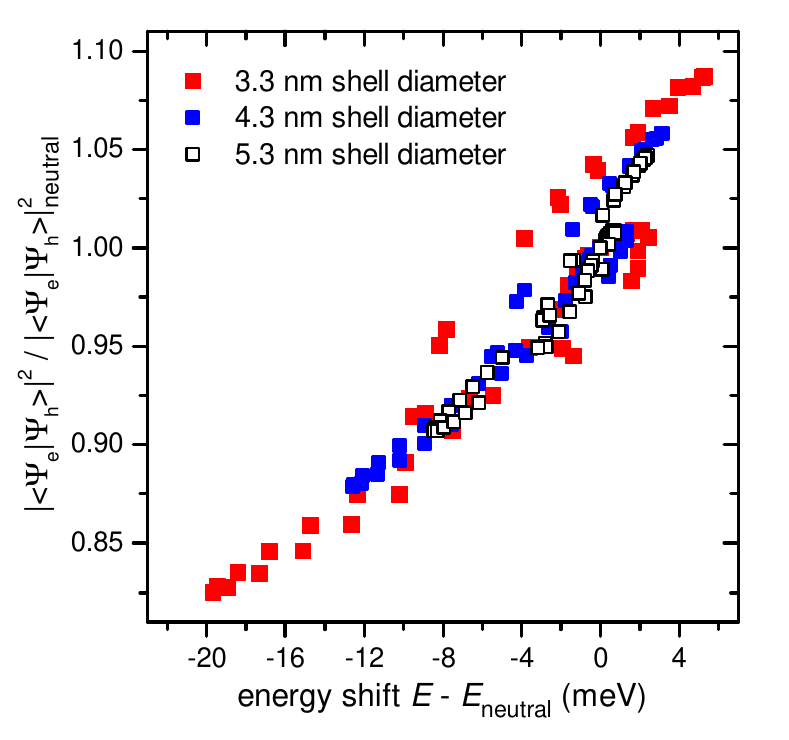}
    \caption{Correlation between the relative squared wave function overlap and the energy shift ---both with respect to the values of the uncharged system--- for DR structures with the same CdSe core size (1.8 nm) and CdS shell length (22 nm), but different shell thicknesses (color coded).}
\label{figure5}
\end{figure}

As explained in the previous section, we use individual surface charges placed at different locations along the DR's $z$-axis [cf.\ Fig.\ \ref{figure4} (a)] to model spectral diffusion. The discussion of Fig.\ \ref{figure4} (b) and (c) revealed that at a certain distance to the CdSe core, the surface charges do not significantly impact the exciton properties.
Thus a variation of the shell length is redundant and not discussed here.
Figure \ref{figure5} shows the correlation between the relative squared wave function overlap and the exciton-energy shift for CdSe/CdS DRs with a fixed core size (1.8~nm) and a set shell length (22~nm) but for three different shell diameters (3.3, 4.3, and 5.3~nm).
The figure is similar to Fig.\ \ref{figure4} (d), but it does not discriminate between the effect of positive or negative individual surface charges.

The general trend of a decreasing squared wave function overlap for red-shifted exciton energies is valid for all shell diameters. Differences occur for the ranges of values for both, overlap and energy: thinner shells lead to larger spreads of the squared wave function overlap and of the exciton energies. 
This can be explained by a stronger interaction between the charge carrier wave functions and migrating surface charges for thinner shells.

The overall slope of the correlation in Fig.\ \ref{figure5} is not significantly influenced by the CdS shell diameter.
In this context it is interesting to note that the slope of the correlation changed when DR structures were modeled either without or with strain contributions, as discussed with respect to Fig.\ \ref{figure4} (d).
This shows that the variation in shell thickness influences the exciton energy and wave function overlap equally, while the strain-induced band-offset change  affects these properties differently.
Decreasing the shell thickness significantly raises the confinement energy contribution in $xy$-direction, while leaving the $z$-confinement unchanged.
As a consequence the delocalization of the charge carrier wave function is also only influenced in the shorter dimensions of the anisotropic geometry.
In comparison, when changing the effective band offset by taking strain into account, the confinement energy is affected in all three dimensions and, because the shell geometry inherently allows for a stronger delocalization in $z$-direction, the squared wave function overlap is affected more significantly for these structures.

\begin{figure}[htbp]
  \includegraphics{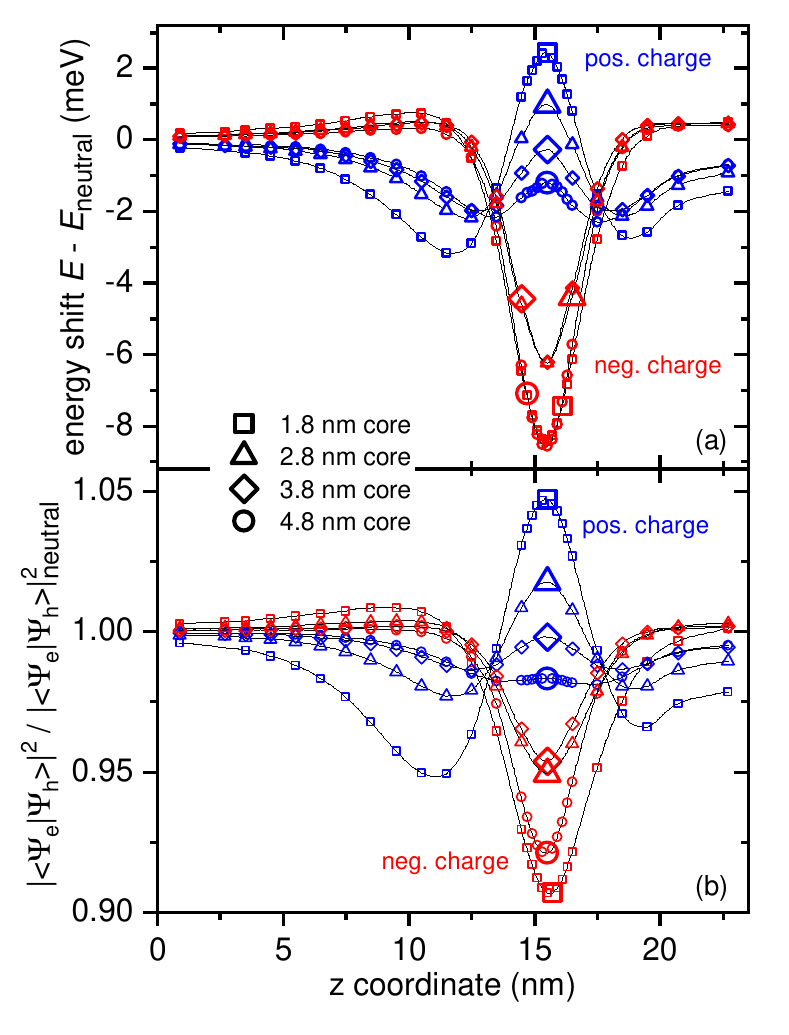}
    \caption{
(a) Surface-charge induced shift of the exciton energy and (b) relative squared wave function overlap in dependence of the surface-charge's $z$-position calculated for DR structures with the same CdS shell dimensions (22 nm length and 5.1 nm thickness) but with different CdSe core sizes (color coded). Reference points for the calculations are the values for the corresponding neutral CdSe/CdS DR structure. In the DR model structures the  CdSe core center was located at $z= 15.33$~nm.}
\label{figure6}
\end{figure}

We now want to discuss the influence of the core size on the spectral diffusion in strained DR model structures. Similar to Fig.\ \ref{figure4} (b) and (c) for unstrained and strained DRs, Fig.\ \ref{figure6} (a) and (b) show changes in exciton energy and squared wave function overlap in dependence on the surface charge position in this case for strained CdSe/CdS DRs only, but for  four different core diameters (1.8 to 4.8~nm, in 1~nm steps). In these model structures, the shell geometry is always the same as for the DRs discussed in Fig.\ \ref{figure4}, i.e. the shell length and diameter have been set to 22~nm and 4.8~nm, respectively.
The curve for the 1.8 nm-sized core corresponds to the calculated data depicted as orange symbols in Fig.\ \ref{figure4} (b) and (c).
The curves retain their general shape also for increasing core sizes, thus, the discussion of the curve shape given with respect to Fig.\ \ref{figure4} (b) and (c) is still valid.

Several interesting details occur in Fig.\ \ref{figure6}.
One of the details is related to the minima in both energy and overlap that occur for a negative surface charge at the position of the core. Here, with increasing core diameter the strength of the minima first decreases (like it does for the maxima in case of a positive surface charge), but, interestingly, for the largest core diameter, the minima in energy and overlap nearly reach the values for the smallest core again. This behavior can be rationalized as follows: In the case of the large core, not only the hole but also the electron has its wave function localized predominantly inside the core, leading to an inherently large squared wave function overlap. Furthermore, a large core comes along with a thin shell, hence, the distance between surface charge and exciton is small.
The Coulomb potential of the surface charge is steep for very small distances. In the case of a thin shell,  a negative surface charge induces a steep Coulomb potential inside the core that attracts and strongly binds the hole. This effectively decreases the energy of the exciton.
Other details, like the negative energy shift and the relative overlap smaller than one for large cores when a positive surface charge is placed at the position of the core, which we have already addressed in Ref.\ \citenum{lohmann_surface_2017}, can be explained in a similar but rather complex way as above. 
The complexity of the explanations is a consequence of the counteracting effects in passivation for decreasing core sizes and increasing shell thicknesses. 
Furthermore, when comparing individual curves in Fig.\ \ref{figure6} (a) or (b), one has to keep in mind that the excitonic properties of a certain DR geometry is displayed relative to the corresponding uncharged structure. Thus the points of reference for each curve are different.

Summarizing the effect of the geometry on the spectral diffusion, we find that for a certain CdSe core size the CdS shell thickness defines the ranges of exciton energy and squared wave function overlap.
Thinner CdS shells lead to a larger variation of the exciton properties.
In case of a constant CdS shell diameter, surface charges influence the exciton properties strongest for small cores.

\section{Summary}

We have modeled the effects of strain on the optical properties of CdSe/CdS core/shell DRs. Questions that have been addressed include the issue of a possible quasi-type-II band alignment in these (strained) hetero-nanostructures and the impact of strain on the optical stability of DR emitters, in particular on the spectral diffusion.

To incorporate strain effect in the modeling, a valence force field approach has been used. First an atomistic model structure has been defined by assuming a wurtzite CdS rod with bulk bond parameters, which has the sulphur atoms exchanged by selenium atoms in the spherical volume of the core positioned inside the rod. Assuming bulk bond lengths also for the such defined CdSe core, the whole hetero-nanostructure has been relaxed by minimizing its total energy. 
The resulting strained atomistic DR model reveals a compressed CdSe core and an anisotropically strained CdS shell with its strain characteristics depending on the exact geometry of the DR. The calculated strain in bond lengths has been used to determine the strain-induced volume changes of CdSe and CdS unit cells in the DR. These volume changes have then been converted to changes in the valence and conduction band edge energies, via the deformation potential. 
The spatially varying band edges of the model structure have then been projected on a regular three-dimensional cartesian grid. This projection serves as input for the EMA-based calculations of electron and hole eigenenergies and wave functions.    

Investigating the band alignment in the CdSe/CdS DRs at the core/shell interface, we find that the inclusion of strain effects leads to a decrease of the band-edge offsets. However, even for very small cores, the corresponding decrease is not strong enough to change the 300 meV type-I conduction band offset of the CdSe/CdS heterostructure into a quasi-type-II or even type-II band alignment. 

For unstrained as well as for strain-relaxed DR model structures, the exciton energy nearly linearly decreases with increasing core size of DRs with otherwise identical geometry. At the same time, the wave function overlap of electron and hole is increasing with increasing core size. Differences in the overlap of the unstrained and strained DR model structures are largest for intermediate core sizes. 
In general, the wave function overlap stays considerably large, much larger than one would expect for real quasi-type-II nanoparticles. Nevertheless, it is true that the electron wave function slightly protrudes into the shell of the DR and that this protrusion is enhanced by strain effects.

In an earlier work, we have modeled spectral diffusion in DR structures by migrating surface charges. In particular, Coulombic point charges have been placed at different positions on the the DR surface and their impact on the excitonic properties have been determined using again EMA-based calculation. 
In order to address the strain effects on the spectral diffusion of CdSe/CdS DRs, here, we combine above described strain model with the model of migrating surface charges. 
We find that the inclusion of strain effects into the model of migrating surface charges leads to a larger range in energy shifts as well as squared wave function overlap changes. The correlation of both, the surface-charge induced energy shift and the relative squared wave function overlap is of special interest since it connects to the experimentally accessible correlation between lifetime and energy of the DR fluorescence emission. We find that strain-effects also change the slope in the calculated correlation. 
Interestingly, including the strain into the spectral-diffusion model modifies the theoretical results in a way that they fit better to the experimental data given in Ref.\ \citenum{lohmann_surface_2017}.

Finally, we have analyzed the impact of the CdSe/CdS DR geometry on the spectral diffusion under consideration of strain.  
For a fixed CdSe core size, an increasing CdS shell thickness yields decreasing ranges of the spectral shifting and the squared wave function overlap, while the slope of their correlation is only marginally changed. 
Changing the CdSe core size while keeping the CdS shell diameter fixed reveals that surface charges have the strongest impact on the excitonic properties when the core size is small. 

Our work can help the tailored design of CdSe/CdS DRs with suppressed fluorescence spectral diffusion. 
This is important for their application in light-emitting devices, such as displays or even future single-photon sources, in which spectral diffusion is generally undesirable since it diminishes the emitter quality.
We show that the fluorescence spectral diffusion behavior can be tuned by the core size as well as shell thickness.
Minimization of the range of both energy and decay rates can be achieved by employing thick CdS shells combined with preferably large CdSe cores.
Furthermore the universal approach of our method allows the prediction of spectral diffusion properties and further optical properties not only for CdSe/CdS DRs of different geometries, but also for hetero-nanostructures of entirely different material combinations and arbitrary geometries.
An example of heterostructured nanoparticles with a similar geometry are ZnSe/CdS DRs, but they are expected to have significantly different spectral-diffusion properties. This is because of the inherent type-II band alignment between ZnSe and CdS, which leads to an entirely different  charge-carrier localization in the DRs compared to the CdSe/CdS system. 

As a final remark we want to note that in our current model, the effects of piezoelectric fields have not yet been implemented. These effects are considered to be small for the CdSe/CdS DRs discussed here, but may be more significant for other material systems or nanoparticle geometries. \cite{segarra_piezoelectric_2016}

\section{Acknowledgements}

S.-H.~L. thanks the Fonds der Chemischen Industrie (FCI) for financial support. This work has been supported by the excellence cluster ‘The
Hamburg Centre for Ultrafast Imaging – Structure, Dynamics
and Control of Matter at the Atomic Scale’ of the Deutsche
Forschungsgemeinschaft via EXC1074.


\bibliography{BibStrain}

\newpage
\section{Graphical TOC Entry}
\begin{figure}
  \includegraphics{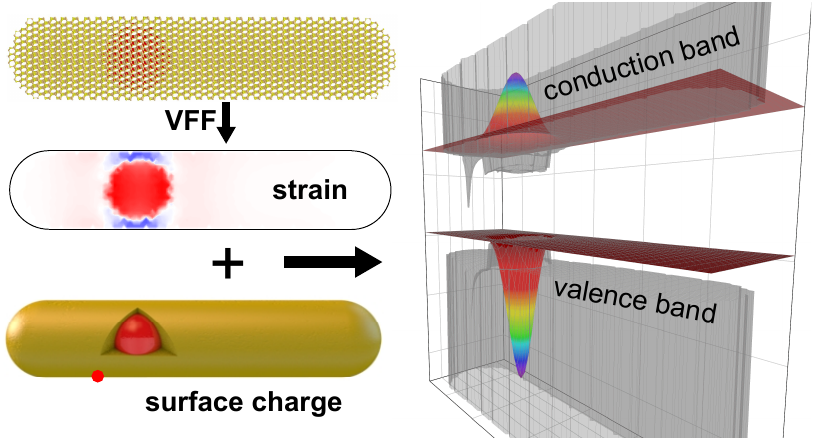}
  \caption*{Table of Contents Graphic}
\end{figure}

\end{document}